\documentclass[final,1p,times]{elsarticle}

\usepackage[fleqn,tbtags]{mathtools}
\usepackage{amssymb}
\usepackage[utf8]{inputenc}
\usepackage[OT4]{fontenc}
\usepackage{hyperref} 
\usepackage{graphicx} 
\usepackage{psfrag}
\usepackage{subfig}

\journal{Physica A}

\begin{document}

\begin{frontmatter}



\title{Gain and loss of esteem, direct reciprocity and Heider balance}

\author[ir,agh1]{Forough Hassanibesheli}  
\author[ir,agh1]{Leila Hedayatifar} 
\author[agh1]{Przemysław Gawroński\corref{cor}}
\ead{gawron@newton.ftj.agh.edu.pl}
\author[agh2]{Maria Stojkow} 
\author[agh2]{Dorota Żuchowska-Skiba} 
\author[agh1]{Krzysztof Kułakowski}

\address[ir]{Department of Physics, Shahid Beheshti University, G. C., Evin, Tehran 19839, Iran}
\address[agh1]{AGH University of Science and Technology, Faculty of Physics and Applied Computer Science, al. Mickiewicza 30, 30-059 Kraków, Poland}
\address[agh2]{AGH University of Science and Technology, Faculty of Humanities, al. Mickiewicza 30, 30-059 Kraków, Poland}

\cortext[cor]{Corresponding author}


\begin{abstract}
The effect of gain and loss of esteem is introduced into the equations of time evolution of social relations, hostile or friendly, in a group of actors. The equations allow for asymmetric relations. We prove that in the presence of this asymmetry, the majority of stable solutions are jammed states, i.e. the Heider balance is not attained there. A phase diagram is constructed with three phases: the jammed phase, the balanced phase with two mutually hostile groups, and the phase of so-called paradise, where all relations are friendly.
\end{abstract}

\end{frontmatter}

\section{Introduction}
Despite the eternal doubts whether social phenomena can be described quantitatively \cite{weber,elias,burawoy}, mathematical modeling of interpersonal relations has long tradition \cite{festinger,heider,coleman,nowak,helbing,friedkin,bonacich}. The idea is tempting: a bottom-up path from understanding to control and predict our own behaviour seems to promise a higher level of human existence. On the other hand, any progress on this path is absorbed by the society in a kind of self-transformation, what makes the object of research even more complex. As scholars belong to the society, an observer cannot be separated from what is observed; this precludes the idea of an objective observation. Yet, for scientists, the latter idea is paradigmatic; their strategy is to conduct research as usual. As a consequence, the hermeneutically oriented multi-branched mainstream is accompanied by a number of works based on agent-based simulations, statistical physics and traditional, positivist sociology. 

Theory of the Heider balance \cite{heider} is one of such mathematical strongholds in the body of social science. Based on the concept of the removal of cognitive dissonance (RCD) \cite{festinger}, it has got a combinatorial formulation in terms of graph theory \cite{harary}. In a nutshell, the concept is as follows: interpersonal relations in a social network are either friendly or hostile. The relations evolve over time as to implement four rules: friend of my friend is my friend, friend of my enemy is my enemy, enemy of my friend is my enemy, enemy of my enemy is my friend. In a final 'balanced' state, the cognitive dissonance is absent \cite{harary,cope}. There, the network is divided into two parts, both internally friendly and mutually hostile. A special case when all relations are friendly (so-called paradise) is also allowed. More recently, Monte-Carlo based discrete algorithms have been worked out to simulate the dynamics of the process of RCD on a social network \cite{antal}. In parallel, a set of deterministic differential equations has been proposed as a model of RCD \cite{kk,pg1,pg2}. This approach has been generalized to include asymmetric relations \cite{gawronski} as well as the mechanism of direct reciprocity \cite{krawczyk}, which was supposed to remove the asymmetry. 

Our aim here is to add yet another mechanism, i.e. an influence of the rate of the change of relations to the relations themselves. This mechanism has been described years ago by Elliot Aronson and termed as the gain and loss of esteem \cite{linder}; see also the description and literature in \cite{aronson}. Briefly, an increase of sympathy $x_{ij}$ of an actor $i$ about another actor $j$ appears to be an independent cause of sympathy $x_{ji}$ of the actor $j$ about the actor $i$. By 'independent' we mean: not coming from $x_{ij}$, but from the time derivative $\frac{dx_{ij}}{dt}$. In summary, both the relation $x_{ij} $ itself and its time derivative influence the relation $x_{ji}$. The efficiencies of these impacts and the rate of RCD play the roles of parameters in our model. We note that the concept of gain and loss of esteem has triggered a scientific discussion which is not finished until now \cite{tognoli,lawrence,lehr,filipowicz,reid}. Among implications, let us mention two: for man-machine cooperation \cite{komatsu} and for evaluations of leaders as dependent on the time evolution of their behaviour (the so-called St. Augustine effect) \cite{allison}. In our opinion, it is worthwhile to try to include the effect in the existing theory of RCD.

Here we are interested in three phases of the system of interpersonal relations: the jammed phase, the balanced phase with two mutually hostile groups, and the phase of so-called paradise, where all relations are friendly. The two latter phases are known from early considerations of the Heider balance in a social network \cite{harary}. The jammed phase is the stationary state of relations, where the Heider balance is not attained. Jammed states have been discussed for the first time for the case of symmetric relations (i.e. $x_{ji}=x_{ij}$) by Tibor Antal and coworkers \cite{antal}. The authors have shown, that this kind of states appear rather rarely, and the probability of reaching such a state decreases with the system size. A similar conclusion has been drawn also for the evolution ruled by the differential equations \cite{strogatz}.

Our goal here is twofold. First, we provide a proof that with asymmetric relations, the number of jammed states is at least $N$ times larger than the number of balanced states, where $N$ is the number of nodes. The conclusion of this proof is that if the jammed phase is possible, it is generic. Second, we construct a phase diagram, with the model parameters as coordinates, where ranges of parameters are identified where the three above states appear.

In the next section we give a proof that for asymmetric relations the majority of stationary states are jammed states. Third section is devoted to the generalized form of the model differential equations which govern RCD, with all discussed mechanisms included. Numerical results on the phase diagram are shown in Section 4. Final remarks close the text.

\section{The jammed states}
In \cite{antal}, a discrete algorithm (the so-called Constrained Triad Dynamics) has been proposed to model RCD. For each pair of nodes $(ij)$ of a fully connected graph, an initial sign $x_{ij}=\pm 1$ (friendly or hostile) is assigned to the link between nodes $i$ and $j$. For this initial configuration, the number of imbalanced triads $(ijk)$ (such that $s_{ij}s_{jk}s_{ki}<0$) is calculated. This number can be seen as an analogue to energy; let us denote it as $U$. The evolution goes as follows. A link is selected randomly. If the change of its sign lowers $U$, the sign is changed; if $U$ increases, the change is withdrawn; if $U$ remains constant, the sign is changed with probability 1/2. Next, another link is selected, and so on. As a consequence, in a local minimum of $U$ the system cannot evolve; the state is either balanced or jammed.

According to \cite{antal}, the minimal network size where jammed states are possible is $9$ nodes. An example of jammed states for $N=9$ is the case of three separate triads, where each link within each triad is positive (friendly) and each link between nodes of different triad is negative (hostile). As shown in \cite{antal}, each modification of a link leads to an increase of the number of unbalanced triads above the minimal value $U=27$. This can be seen easily when we notice that for each unbalanced triangle, each its vertex is in different triad; hence the number of these triangles is $3\times 3\times 3=27$. A simple inspection tells us that a change of any friendly link to hostile state enhances energy to $U=34$, while any opposite change gives $U=28$; hence the configuration is stable.

The same can be shown for the differential equations; for the purpose of this check we can limit the equations to the simplest form
\begin{equation}
\frac{dx_{ij}}{dt}=\Theta(x_{ij})\Theta(1-x_{ij})\sum_k^{N-2}x_{ik}x_{kj}.
\label{simple}
\end{equation}
The condition of stability of a 'saturated' state where for all links $|x_{ij}|\approx 1$ is that $x_{ij}\frac{dx_{ij}}{dt}>0$; either $x_{ij}$ is positive and increases, or it is negative and decreases. For the above jammed state, all positive links are within a triad, and each negative link is between different triads. Provided that the product of the $\Theta$ functions is equal to one, the r.h.s. of Eq. (\ref{simple}) is equal to $+7$ for positive links, and to $-1$ for negative links; hence this configuration is stable again. This equivalence of the results for two different frames of calculation, discrete and continuous, comes from the fact that in both frames the state of one link can be changed independently on other links. This is not so, if we modify a whole triangle, as in the Local Triad Dynamics \cite{antal}; there, jammed states do not appear.

Now let us consider jammed states in asymetric matrix of relations, where the condition $x_{ji}=x_{ij}$ does not hold. In \cite{krawczyk}, a series of such states has been identified by a numerical check. Here we show that all of them can be captured in the following scheme. Take one node (say $i$-th one), and set all links $x_{ij}=1$, and $x_{ji}=-1$ (or the opposite). As the $i$-th node can be selected in $N$ ways, there are $2N$ such states. Now pick up one node again and divide the remaining network of $N-1$ nodes into two non-empty, internally friendly and mutually hostile sets, with nodes marked by $j$ and $k$, respectively. Without the preselected $i$-th node, the system is balanced. The division can be performed in $2^{N-2}-1$ ways. The relations of the $i$-th node with the others are as follows: either $x_{ij}=1$, $x_{ki}=1$, $x_{ji}=-1$, $x_{ik}=-1$, or $x_{ij}=-1$, $x_{ki}=-1$, $x_{ji}=1$, $x_{ik}=1$. These two options mean that we have $2N(2^{N-2}-1)$ jammed states. In summary, we get $2N+2N(2^{N-2}-1)=N2^{N-1}$ jammed states.

\begin{figure}[!hptb]
\begin{center}
\includegraphics[width=.69\columnwidth, angle=0]{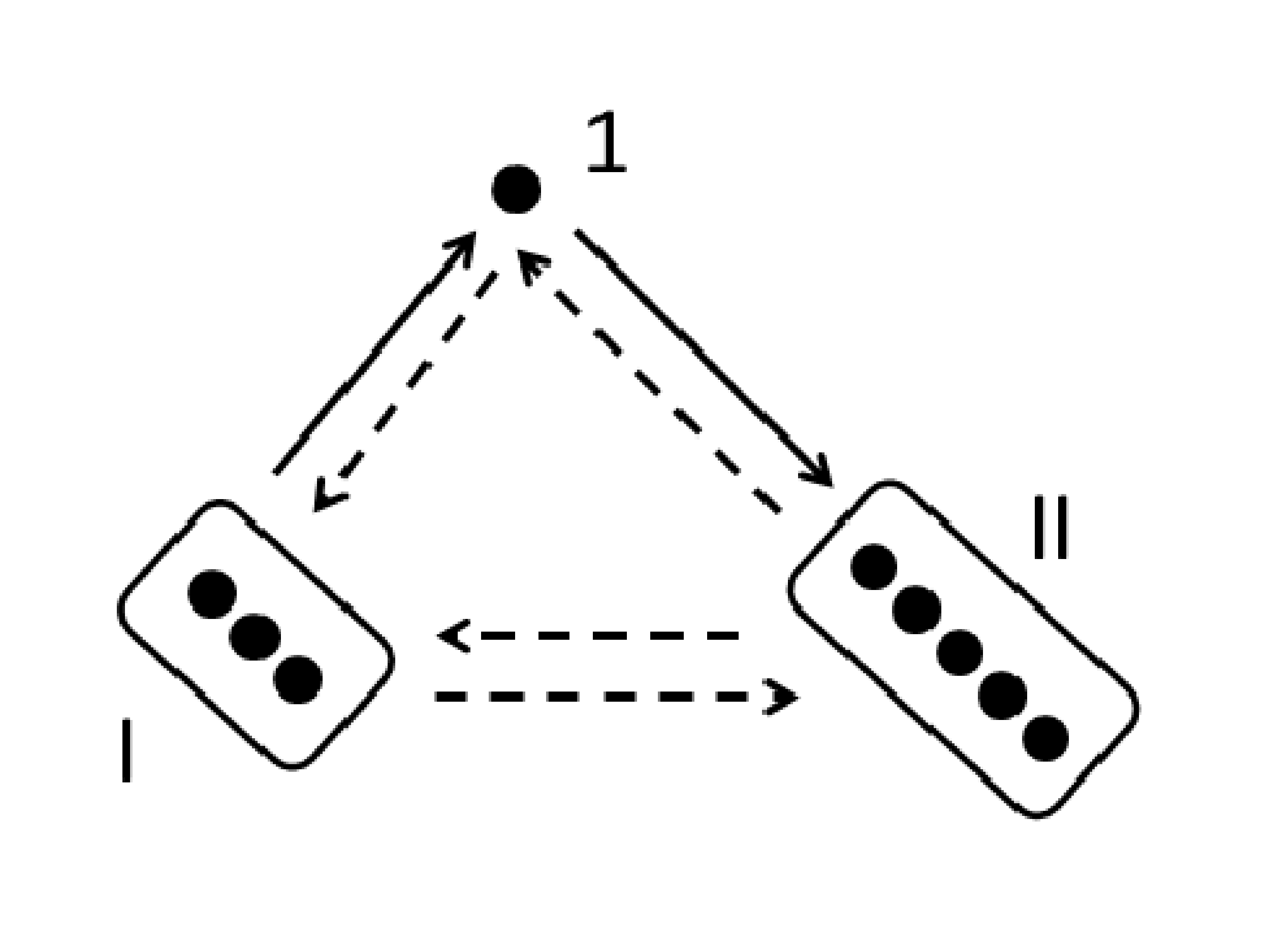}
\caption{The jammed configuration composed of two internally friendly and mutually hostile sets of nodes I, II and one node 1 with asymmetric relations. A continuous arrow from $i$ to $j$ means that $x_{ij}$ is positive (friendly). A broken arrow means that the relation is negative (hostile).}
\label{fig1}
\end{center}
\end{figure}

The stability of these jammed states can be verified analytically. Consider a configuration of $N_1$ nodes in part I of network, $N_2$ nodes in part II, and one pre-selected node $i=1$; then, $N=N_1+N_2+1$. As above, the system is balanced ($N_1$ against $N_2$) except node $1$. Suppose that for $j\in$ I and $k\in$ II we have $x_{j1}=x_{1k}=1$ and $x_{1j}=x_{k1}=-1$, as in Fig. \ref{fig1}. Again, the stability can be verified by inspection of the r.h.s. of Eq. (\ref{simple}) for all links. First, each internal links within I is positive. Its time derivative is $N_1-2$ (through other nodes in I), plus $N_2$ (through nodes in II), minus one (through 1); the sum is positive. (Writing 'through node $i$', we mean the contribution $x_{ji}x_{ik}$ to the time derivative of $x_{jk}$.) The same applies to links within II. For $x_{j1}$ (positive), its time derivative is $N_1-1$ (through nodes in I), plus $N_2$ (through nodes in II), what is positive. For $x_{1k}$ (positive), its time derivative is $N_2-1$ (through other nodes of II), plus $N_1$ (through nodes in I, the product of two negative links in each time), what is positive. In the same way, for $x_{1j}$ and $x_{k1}$ (both negative) we get $1-N_1-N_2$, what is negative. 

On the other hand, there is only $2^{N-1}$ balanced states, including the paradise state of all relations friendly. Therefore, the number of jammed states is at least $N$ times larger, than the number of balanced states. We write 'at least' because the numerical identification of jammed states has been done for $N=5$; it is likely, than for larger systems we would observe even more complex configurations. Yet, the numerical time increases with the number of configurations, which is $2^{N^2-N}$.

\section{Equation of motion with gain and lose of esteem}
With all three mechanisms built into the equation, the time derivative of the relation $x_{ij}$ is
\begin{equation}
 \frac{dx_{ij}}{dt}=\left( 1-x_{ij}^2 \right) \left[ \alpha \left(x_{ji}-x_{ij} \right)+\frac{1-\alpha}{N-2}\sum_k^{N-2}x_{ik}x_{kj}+\gamma\frac{dx_{ji}}{dt} \right]
 \label{main}
\end{equation}
Here the coefficient $\alpha$ is a measure of intensity of the process of evening the relations between $i$ and $j$ out because of direct reciprocity. Accordingly, the coefficient $1-\alpha$ is a measure of intensity of RCD. The former mechanism has been introduced in \cite{krawczyk}; for initially asymmetric relations, its appearance is a necessary condition for the Heider balance. Varying $\alpha$, we control the mutual rate of the two processes. However, in stationary states the time derivatives of the relations $x_{ij}$ are equal to zero; therefore the third coefficient $\gamma$, which is a measure of importance of the dynamics, is kept independent on $\alpha$. The prefactor $1-x_{ij}^2$ plays the same role as the product of $\Theta$ functions in Eq. \ref{simple}, i.e. to keep the relations in the prescribed range $(-1,1)$. For numerical calculations, this prefactor is more feasible.

It is not convenient to have time derivatives on the r.h.s. of an equation. Then, after a short algebra, we get
\begin{equation}
\begin{multlined}
\frac{dx_{ij}}{dt} = \frac{(1-x_{ij}^2)}{1-\gamma^2 (1-x_{ij}^2)(1-x_{ji}^2)} \left[ \vphantom{\frac{1-\alpha}{N-2} \sum_k^{N-2}} \alpha \left( 1-\gamma \left( 1-x_{ji}^2 \right) \right) \left( x_{ji}-x_{ij} \right) \right. \\
\left. + \frac{1-\alpha}{N-2} \sum_k^{N-2} \left[x_{ik}x_{kj}+\gamma \left(1-x_{ji}^2 \right)x_{jk}x_{ki} \right] \right]
\end{multlined}
\label{main2}
\end{equation}


It is easy to notice that close to the stationary state, where ($|x_{ij}|\approx 1$), the equation reduces to the case where $\gamma=0$. Yet, as we have found already in \cite{krawczyk}, the action of the mechanism of direct reciprocity is to reduce the asymmetry of the relations before the system is trapped in a jammed state. If the coefficient $\alpha$ is too small, the process is too slow and the Heider balance is not attained. Therefore what is meaningful for the final state is the mutual rate of the processes. Below we present the results of the numerical solution of the system of equations which display the time evolution of $x_{ij}$. How the mechanism of the gain and loss of esteem modifies the action of the direct reciprocity?

\section{Numerical results}
To solve this issue, a series of numerical solutions of Eqs. (\ref{main2}) have been performed, for different values of the parameters $\alpha$ and $\gamma$ and different probability distributions of the initial values of the relations $x_{ij}$. As a rule, the distributions are uniform and different from zero in the range $(p-0.3,p+0.3)$; then, $p$ is yet another parameter. The width of the distribution is chosen to be $0.6$ for all calculations. If the width is too small, the effect of asymmetry is reduced, and too large values do not allow to modify $p$ because the values of $x_{ij}$ are limited to the range $(-1,+1)$.

For each set of trajectories $x_{ij}(t)$, $i,j=1,...,N$, we intend to qualify the final state as one of three phases: the paradise state, the balanced state and the jammed state. As a rule, after a predictable number of time steps the system attained the saturated state where all relations $x_{ij}$ were close to $\pm 1$. Exceptions from this rule have been found to be  very rare and almost all of them happened for $\alpha=0$. Numerical inspection allowed to classify these exceptions as examples of states which varied periodically in time, as identified in \cite{gawronski}. These cases have been eliminated from the statistics.

The state of paradise has been identified from the condition that all relations $x_{ij}$ are $+1$. Similarly, the criterion of the balanced state is that the number of unbalanced triads ($x_{ij}x_{ik}x_{kj}<0$) is zero. If neither of these conditions is fulfilled, the state is classified as a jammed state. In all but one cases, the jammed character of the state was equivalent to a clear asymmetry of the relations, i.e. $x_{ij}\ne x_{ji}$. The very rare exception was that the symmetric relations happened to appear simultaneously with the non-zero number of the unbalanced triads; such states have been discussed in \cite{antal}. The case of three separated triads, found in \cite{antal}, is thoroughly discussed in our Section 2.

Most of our simulations have been performed for $N=20$ nodes, what is equivalent to 380 equations of motion. This choice is motivated by our observation, that the accuracy of numerical solution (limited by the length of the time steps) should be high enough to prevent a jump from one final state of all links to another. This condition in turn makes the time of calculation quite long, if the number of nodes $N$ is about 30. Below $N=20$, the above jumps have not been observed.

Our aim here is the phase diagram, i.e. the ranges of parameters where each of the three phases appear. Here we work with three parameters: $p$, $\alpha$ and $\gamma$. We proceed as follows: for a fixed value of all parameters, we perform a series of $K=200$ simulations and we count, how many times each of three phases appear. The boundary between two phases is set for the set of parameters where each of the two phases appears with the same frequency. The values of $\alpha_c$ are identified from a linear fit between 30 and 70 percent of appearance of a given phase.

\begin{figure}
\begin{center}
\includegraphics[width=.69\columnwidth, angle=0]{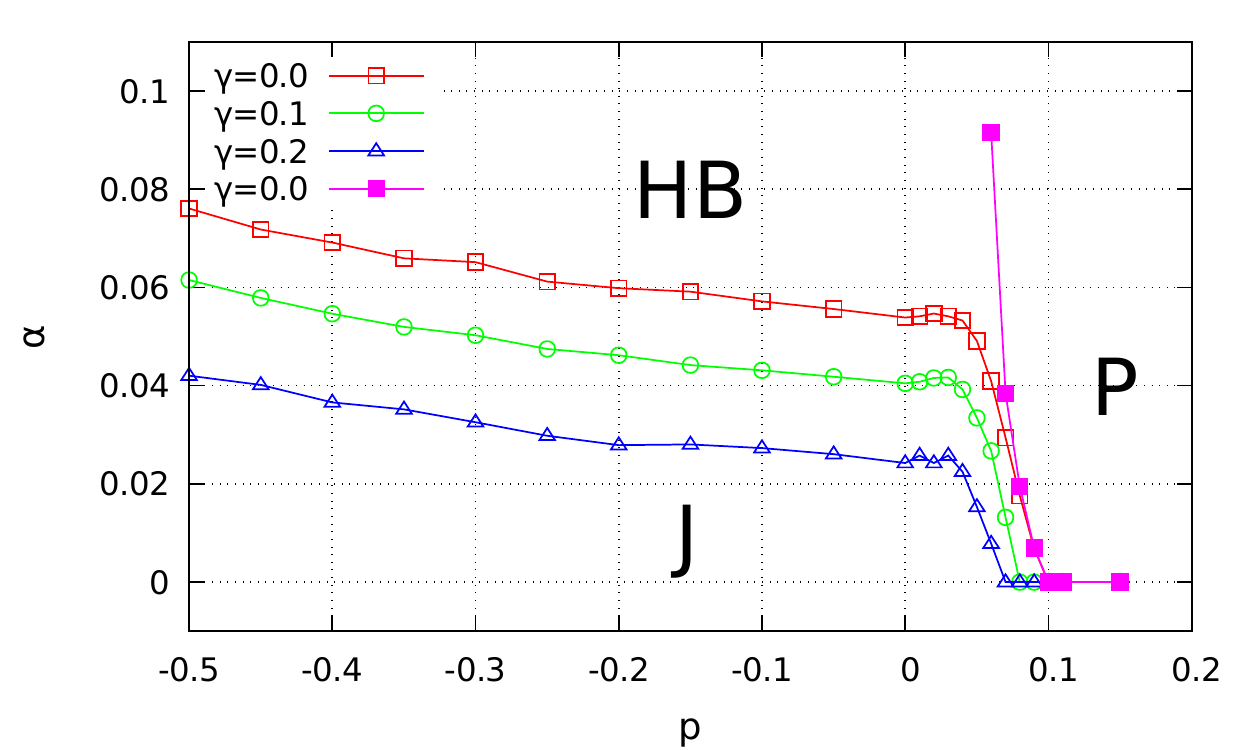}
\includegraphics[width=.69\columnwidth, angle=0]{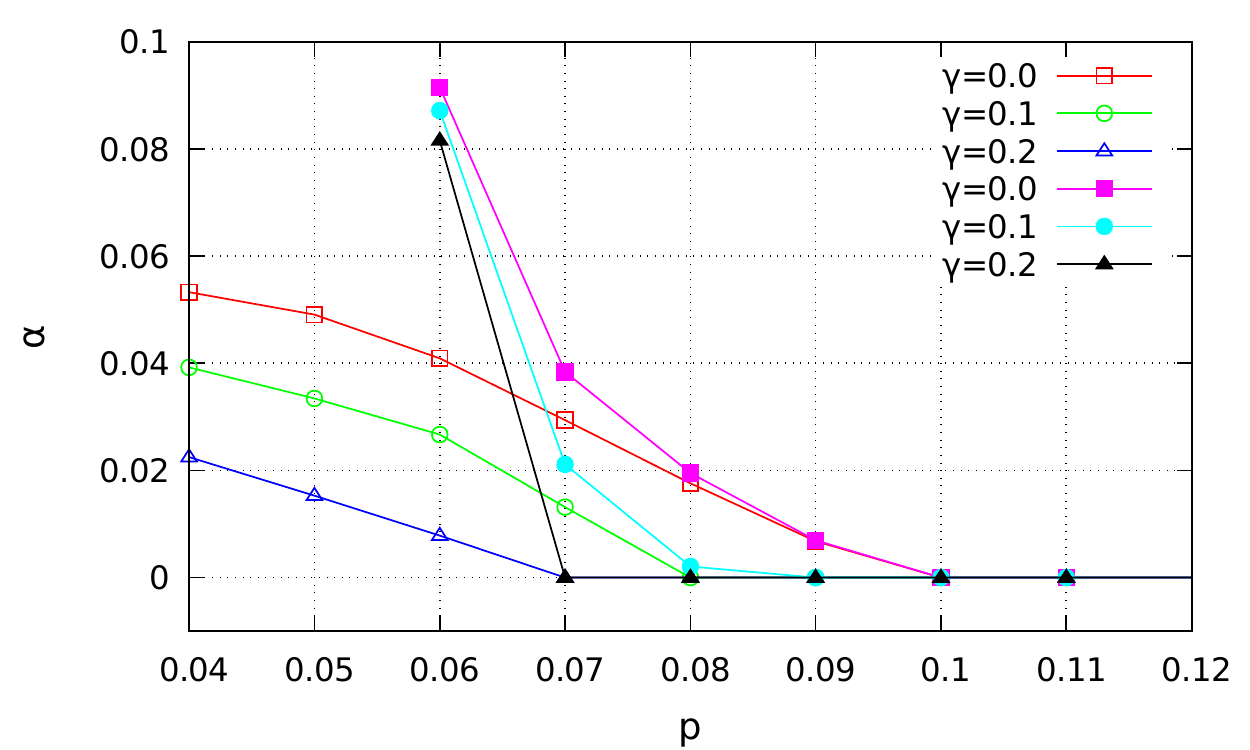}
\caption{The phase diagram on the plane $(p,\alpha)$, with the phases J, HB, P (jammed, balanced, paradise) (upper plot), and an inset (lower plot). The lines are for $\gamma=$ 0.0, 0.1 and 0.2.}
\label{f2}
\end{center}
\end{figure}

Numerical results of these investigations are shown in Fig. (\ref{f2}). There are three plots of $\alpha_c(p)$ for three different values of $\gamma$. The value $\alpha _c$ is the critical value; for $\alpha > \alpha _c$, the jammed phase is more rare than the alternative balanced state or the paradise state. We see that for $p$ larger than about 0.06, the paradise phase appears for all values of $\alpha$; this means that the evolution drives all relations to be friendly ( equal $+1$). Of course, asymmetry vanishes there, even if the symmetrizing mechanism of the direct reciprocity is not present for $\alpha=0$. Below $p$ close to $0.6$, the obtained line $\alpha _c (p)$ is the boundary between the jammed phase ($\alpha < \alpha _c$) and the balanced phase ($\alpha > \alpha _c$). An important result is that when the coefficient $\gamma$ increases, $\alpha _c$ decreases. A less distinct effect is that the transition to the paradise phase is slightly shifted towards smaller $p$, when $\gamma$ increases.

\section{Discussion}
The proof that the majority of stationary states are unbalanced (jammed), given in Section 2, highlights the role of asymmetry of the relations. Recall that in the case of symmetric relations, jammed states are not generic; this is so both for discrete \cite{antal} and continuous \cite{strogatz} dynamics. Basically however, experimentally collected data on relations are not symmetric, as discussed in \cite{carley}; a recent example has been reported in \cite{krawczyk}. Exceptions from this rule can be induced by an experimental method, as when the intensity of communication in pairs is measured \cite{zachary}. A jammed state means that the Heider balance is not attained; we conclude that this is so in most stationary states of network of social relations. As discussed in \cite{antal2}, in some circumstances lack of the balance can prevent conflicts.

There are two aspects of the phase diagram, described above: the mathematical and the sociological one. Let us start from mathematics. Below the critical value of $\alpha$, the mechanism driven by direct reciprocity is not efficient and the system is stuck in a jammed state. What matters is the ratio of efficiencies of both mechanisms, the one proportional to $\alpha$ (direct reciprocity) and the one proportional to $1-\alpha$ (RCD). In Eq. (\ref{main}), the term proportional to $\gamma$ produces another mechanism of evening the asymmetric relations out, because an increase of $x_{ji}$ is correlated with an increase of $x_{ji}$. In the presence of this term, the asymmetry is removed by two mechanisms and not only by one. Therefore, to gain the balance, a smaller value of $\alpha$ is necessary; hence, the critical value of $\alpha$ decreases with $\gamma$, as we see in Fig. (\ref{f2}).

On the sociological side, the results become part of current discussions on the role of cognitive processes in shaping interpersonal relations (\cite{reid,montoya,singh} and references therein). Expressed opinions allow to evaluate the level of friendliness toward us \cite{montoya}. Results of experimental research show that a positive evaluation of a person triggers a reverse sympathy stronger than similarity of attitudes \cite{singh}. In \cite{reid}, the roles of attitude similarity and attitude alignment have been found to act separately and interact; similarity predicted attraction only if attitude alignment was absent. In our formulation RCD (term proportional to $1-\alpha$), positive evaluation (term proportional to $\alpha$) and attitude alignment (term proportional to $\gamma$) act separately. Our phase diagram indicates that they also interact; the decrease of $\alpha _c$ with $\gamma$ means, that if the attitude alignment is strong enough, the positive evaluation is not needed to restore the balanced state.

\section*{Acknowledgements}
One of the authors (K.K.) is grateful to Katarzyna Krawczyk-Szczepanek for kind and helpful comments. The work of P.G. was partially supported by the PL-Grid Infrastructure.

\section*{References}

\end{document}